\documentclass{optica-article}

\journal{opticajournal} % for journals or Optica Open

\articletype{Research Article}

\usepackage{lineno}
\usepackage{soul,color,xcolor}

\begin{document}

\title{A 1.8-µm pitch, 47-ps jitter SPAD Array in 130nm SiGe BiCMOS Process}

\author{Feng Liu,\authormark{1,* } and Edoardo Charbon\authormark{1,**}}

\address{\authormark{1}Advanced Quantum Architecture Laboratory (AQUA), Ecole Polytechnique Fédérale de Lausanne (EPFL), 2002 Neuchâtel, Switzerland}

\email{\authormark{*}liufengthu@gmail.com} 
\email{\authormark{**}edoardo.charbon@epfl.ch} %% email address is required; see note below about the corresponding author designation

% use {asbstract*} to suppress the copyright line. Copyright information will be added in production

\begin{abstract*} 
We introduce the world’s first SPAD family design in 130\,nm SiGe BiCMOS process. At 1.8\,$\mu$m, we achieved the smallest pitch on record thanks to guard-ring sharing techniques, while keeping a relatively high fill factor of 24.2\%. 4$\times$4 SPAD arrays with two parallel selective readout circuits were designed to explore crosstalk and scalability. The SPAD family has a minimum breakdown voltage of 11\,V, a maximum PDP of 40.6\% and a typical timing jitter of 47\,ps FWHM. The development of silicon SPADs in SiGe process paves the way to Ge-on-Si SPADs for SWIR applications, and to cryogenic optical interfaces for quantum applications.

\end{abstract*}

%%%%%%%%%%%%%%%%%%%%%%%%%%  body  %%%%%%%%%%%%%%%%%%%%%%%%%%
\section{Introduction}
Single-photon avalanche diodes (SPADs) have received increasing attention in recent years thanks to their capability to perform high-speed photon counting and time-of-flight (TOF) imaging at picosecond timing resolution, while a Ge-on-Si SPAD is a good candidate for short wave infrared (SWIR) applications.
The advantage of operating in the SWIR is that the solar background is significantly reduced, and the particle scattering is also decreased. With a higher eye-safety threshold in the SWIR than the visible, much higher optical power source can be used to improve the signal-to-noise ratio.
There are at least two approaches to design the Ge-on-Si SPADs.  The first is to design the high performance SPAD on a dedicated platform \cite{vines2019high} , however circuits are generally not scalable and hard to miniaturize. Alternatively, one can add a Ge layer on top of a CMOS SPAD  \cite{Benhammou2023}.

In this paper, we propose a new SPAD built on a SiGe BiCMOS process, to the best of our knowledge the world’s first. This approach could open the way to a new generation of Ge SPADs in standard technologies capable of achieving higher speed and low jitter. 

Our SPAD is also highly miniaturized owing to well-sharing  \cite{henderson20103,ziyang20173mum} and guard-ring sharing techniques \cite{morimoto2020high}. Further techniques, such as 3D stacking  \cite{al2016backside} and on-chip microlenses \cite{shimada2022spad,jogi2023}, could further improve fill factor, photon detection probability (PDP), and pitch.
The SPAD family proposed in this paper reaches a minimum pitch of 1.8\,$\mu$m, a maximum PDP of 40.6\%, a fill factor of 24.2\%, and a single-photon time resolution (SPTR) of 47\,ps FWHM. The technology supports low breakdown voltages of sub-15\,V, making it suitable for advanced BiCMOS processes and compact image sensors for scientific, industrial, and consumer applications.

\section{SPAD and Chip Structure}

\subsection{SPAD Arrays}
\begin{figure}[ht!]
\centering\includegraphics[width=14cm]{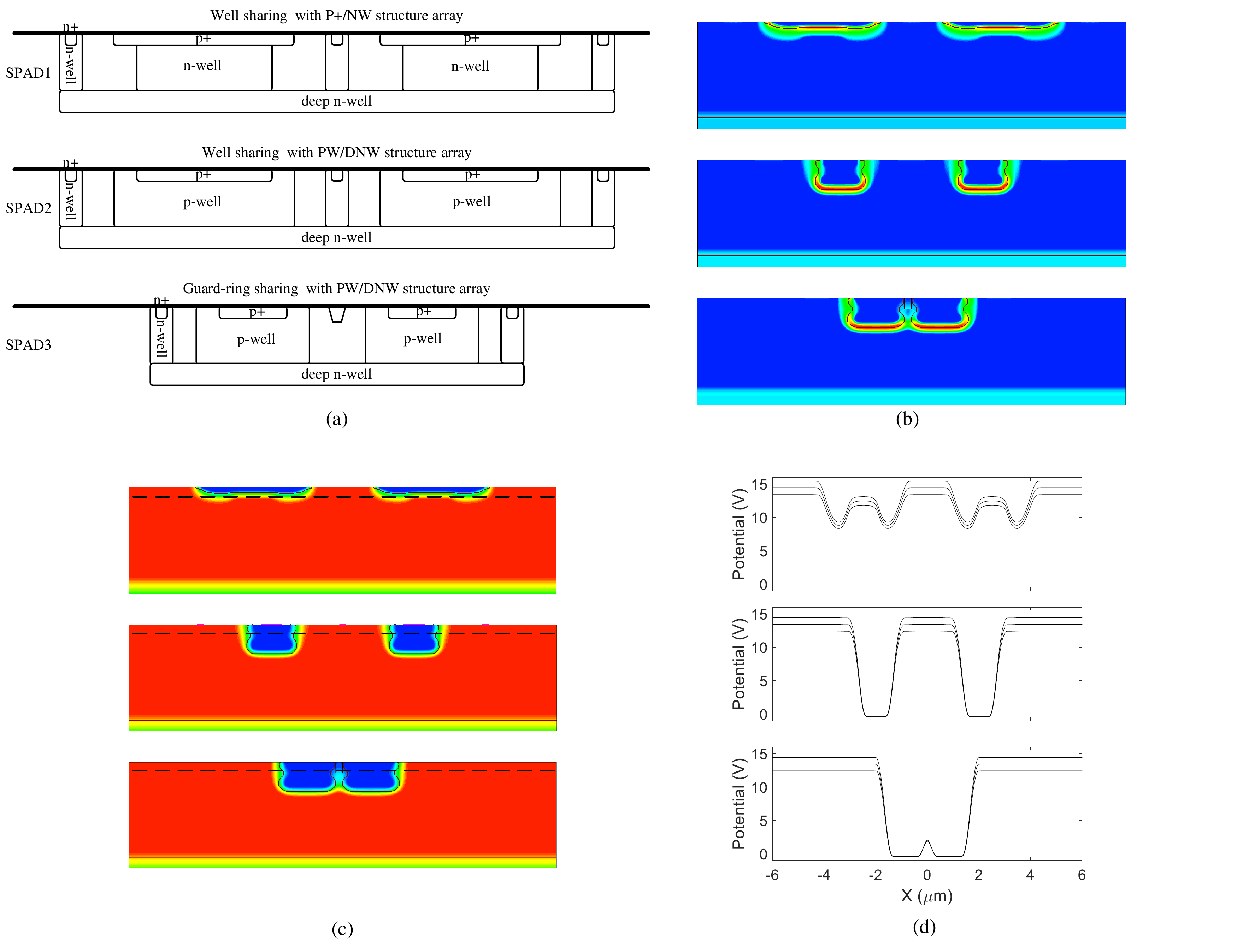}
\caption{Comparison of the SPAD family. (a) Cross-sectional view; (b) 2D view of the simulated electric field distributions; (c) 2D view of the simulated electrostatic potential; (d) 1D potential profile on the dark dashed lines. All the SPADs were designed with a circular drawn active area. The dimensions of the simulated SPAD family are in arbitrary units.}\label{structure}
\end{figure}

Fig.~\ref{structure}(a) shows the cross-section of the SPAD family. SPAD1 is designed with p+/nw structure with n-well sharing technique, SPAD2 is designed with pw/dnw structure with n-well sharing technique. Compared to the standard n-well sharing array, SPAD3 is designed with pw/dnw structure with guard-ring sharing technique, where the pixel is virtually isolated with a shallow trench isolation (STI) [5]. In order to have a systematic performance comparison of the SPAD structure and sharing techniques, SPAD1 and SPAD2 have a similar drawn active diameter with n-well sharing technique to compare the performance of the shallow and deep junction, while SPAD3 has a much smaller SPAD pitch with guard-ring sharing technique to achieve the smallest pitch.
SPAD1 has relatively conservative parameters, i.e. a drawn active diameter of 5.0\,$\mu$m, a pitch of 11.3\,$\mu$m, and a fill factor of 15.3\%, while already using n-well sharing techniques. Thanks to virtual retrograde guard-ring techniques, SPAD2 has a similar drawn active diameter of 5.2\,$\mu$m, and a smaller pitch of 8.3\,$\mu$m, with a fill factor of 30.9\%. Next, in SPAD3, we pushed the pitch scaling to 1.8\,$\mu$m thanks to extensive use of guard-ring sharing techniques, while keeping a relatively high fill factor of 24.2\%. To the best of our knowledge, this is the smallest pitch ever achieved in a SPAD. Fig.~\ref{structure}(b) shows the electric field distribution of a 2D simulation for all proposed SPADs. Note the independent avalanche regions in close proximity in SPAD3. Fig.~\ref{structure}(c) shows the simulated electrostatic potential distribution. Potential profiles are shown in Fig.~\ref{structure}(d) on dark dashed lines. Compared to SPAD1 and SPAD2, SPAD3 has a lower potential barrier height, this was achieved through careful guard-ring structure design, so as to decrease punch-through effects.

\subsection{Pixel Circuitry}
\begin{figure}[ht!]
\centering\includegraphics[width=11cm]{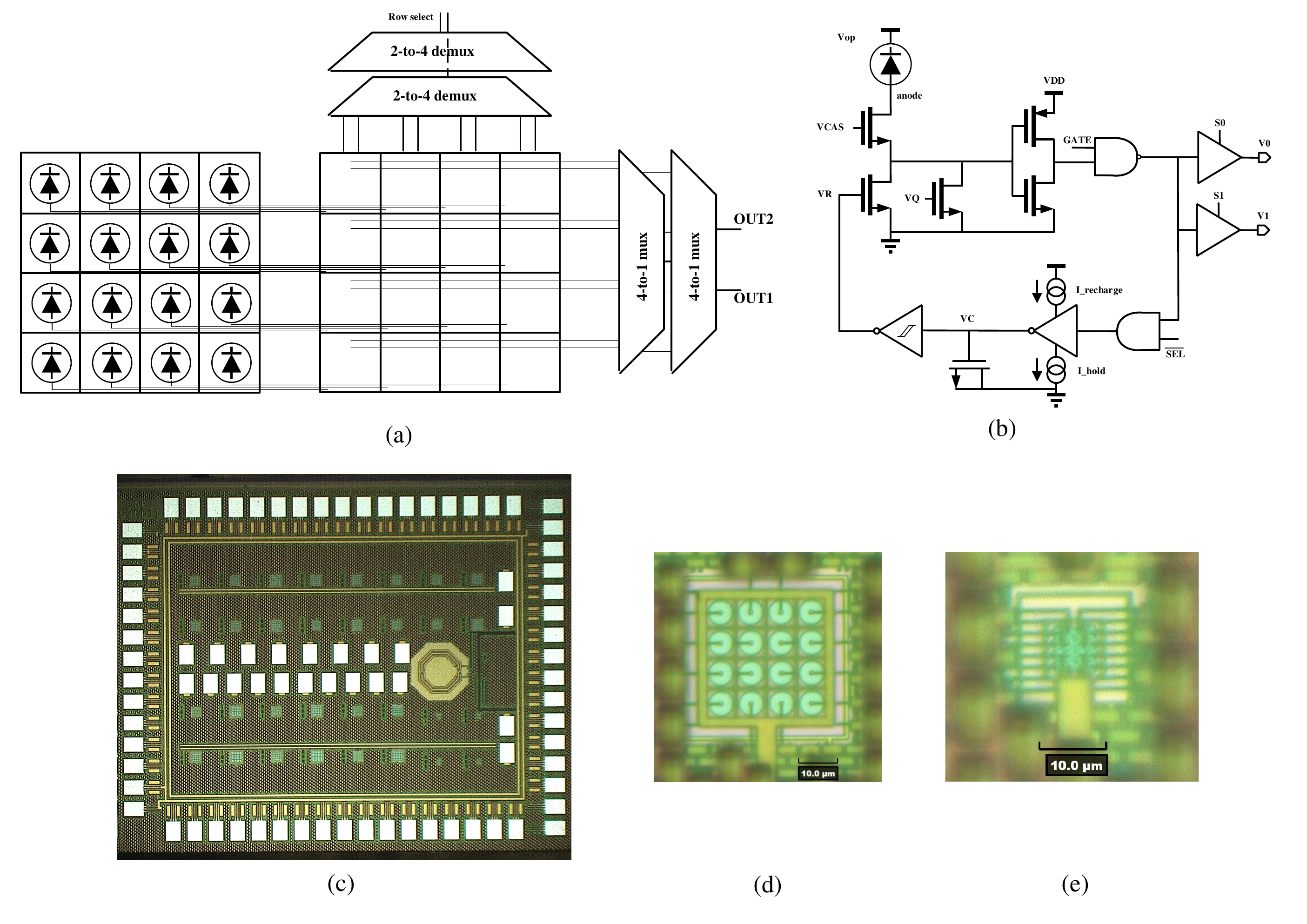}
\caption{ (a) Circuit diagram of one 4×4 SPAD array; (b) Pixel schematic in the 4×4 SPAD array readout. (c) Micrograph of the whole chip with the size of 2.0\,mm × 1.5\,mm. Micrograph of SPAD2 (d) and SPAD3 array (e).}\label{circuit}
\end{figure}

4×4 SPAD arrays with selective output were designed in monolithic 130nm SiGe BiCMOS process. Fig.~\ref{circuit}(a)  shows the circuit diagram of one 4×4 SPAD array, the 16 SPADs share one cathode, while the anode terminal of each SPAD is connected to an array of pixel circuitry. Using demultiplexers and multiplexers, 2 arbitrary SPADs in the array of interest can be streamed out. Fig.~\ref{circuit}(b) shows the pixel diagram used in the 4×4 SPAD array. The front-end circuit consists of passive quenching, and optional active recharging\cite{gramuglia2021engineering}.The estimated quenching resistor is approximately 200 k$\Omega$. The results are based on passive quenching and active recharging, with a dead time of 5\,ns. In Fig.~\ref{circuit}(c), the micrograph of the chip with the size of 2.0\,mm × 1.5\,mm is shown. It includes 32 SPAD arrays with different SPAD structures and dimensions. Fig.~\ref{circuit}(d)-(e) shows the micrograph of SPAD2 and SPAD3 array. All the dummy metals over the SPAD sensitive area were removed by design.

\begin{figure}[ht!]
\centering\includegraphics[width=11cm]{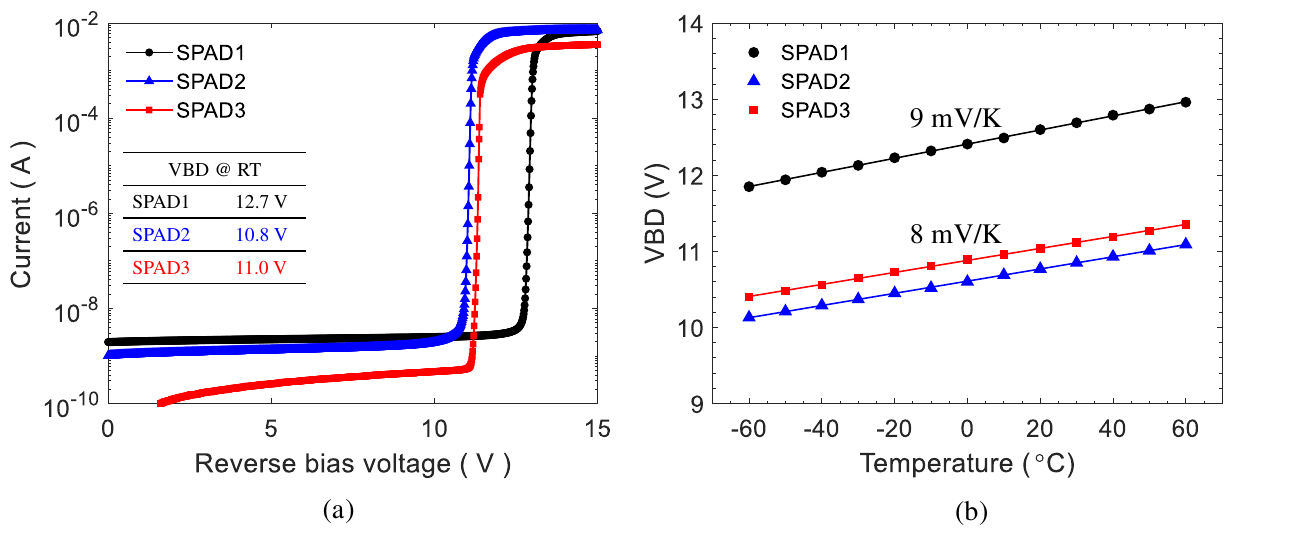}
\caption{ (a) Current-voltage characteristics with illumination at room temperature. (b) Breakdown voltage as a function of temperature of the SPAD family. The breakdown voltage is closely related to the doping concentration of the P-N junction, not the junction depth, directly.} \label{iv}
\end{figure} 

\section{Result and Discussion}

\subsection{I-V Curve}
The static current-voltage curves of the SPAD family with illumination are shown in Fig.~\ref{iv}(a). The results were measured with a high bias voltage condition (VQ\,=\,1.2\,V, VCAS\,=\,3.3\,V). Under the same light intensity, SPAD3 shows a lower current value. The extracted breakdown voltage is 12.7\,V, 10.8\,V, and 11.0\,V, respectively. Fig.~\ref{iv}(b) shows the breakdown voltage as a function of temperature from -60\,ºC to 60\,ºC. The extracted temperature coefficients are 9\,mV/K, 8\,mV/K, and 8\,mV/K, respectively.

\subsection{Dark Count Rate}
\begin{figure}[ht!]
\centering\includegraphics[width=11cm]{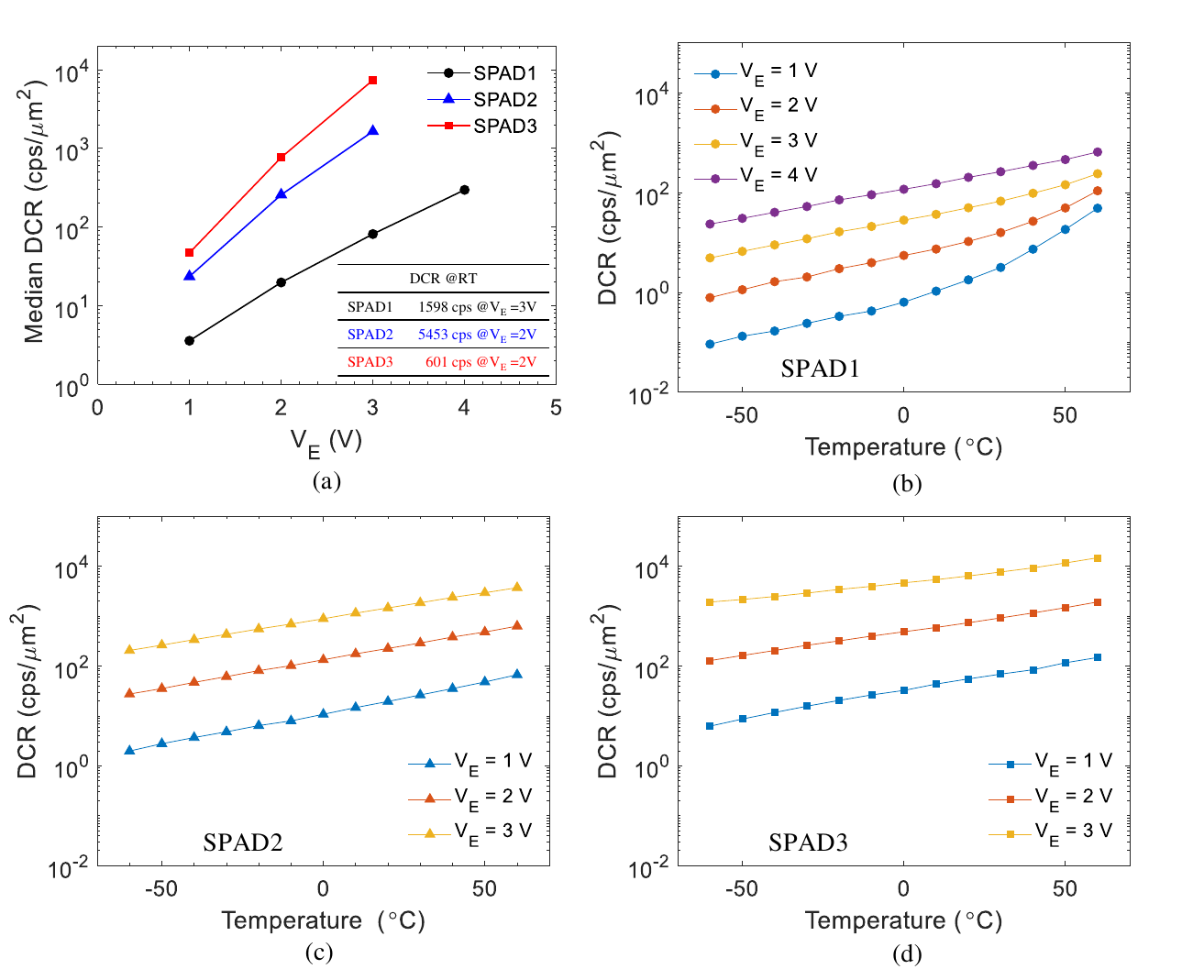}
\caption{ (a) Median DCR as a function of excess bias voltage of SPAD1, SPAD2, and SPAD3. The data is obtained from 16 identical samples. (b) Normalized DCR as a function of temperature with different excess bias voltage of SPAD1 (b), SPAD2 (c), and SPAD3 (c). }\label{dcr}

\end{figure}

Dark count rate (DCR) was measured in 16 samples for all SPADs; Fig.~\ref{dcr}(a) shows the median DCR as a function of excess bias voltage at 25\,ºC. SPAD1 shows the best DCR performance, with a median DCR of 388\,cps at 2\,V and 1598\,cps at 3\,V excess bias voltage. The median DCR of SPAD2 is 5453\,cps and that of SPAD3 is 601\,cps at 2\,V excess bias voltage. Due to crosstalk, SPAD3 shows a worse DCR performance than SPAD2 when we comparing the normalized DCR at the same excess bias voltage. The DCR of the SPAD family as a function of temperature is shown in Fig.~\ref{dcr}(b)-(d). Thermally generated carriers and band-to-band tunneling are the primary contributors to the DCR. In SPAD1, thermal generation dominates at low excess bias voltage and high temperatures, while band-to-band tunneling becomes dominant at high excess bias voltages. In contrast, for in SPAD2 and SPAD3, band-to-band tunneling dominates across all excess bias voltages and temperatures. This effect is primarily due to the high doping concentration.

\begin{figure}[ht!]
\centering\includegraphics[width=11cm]{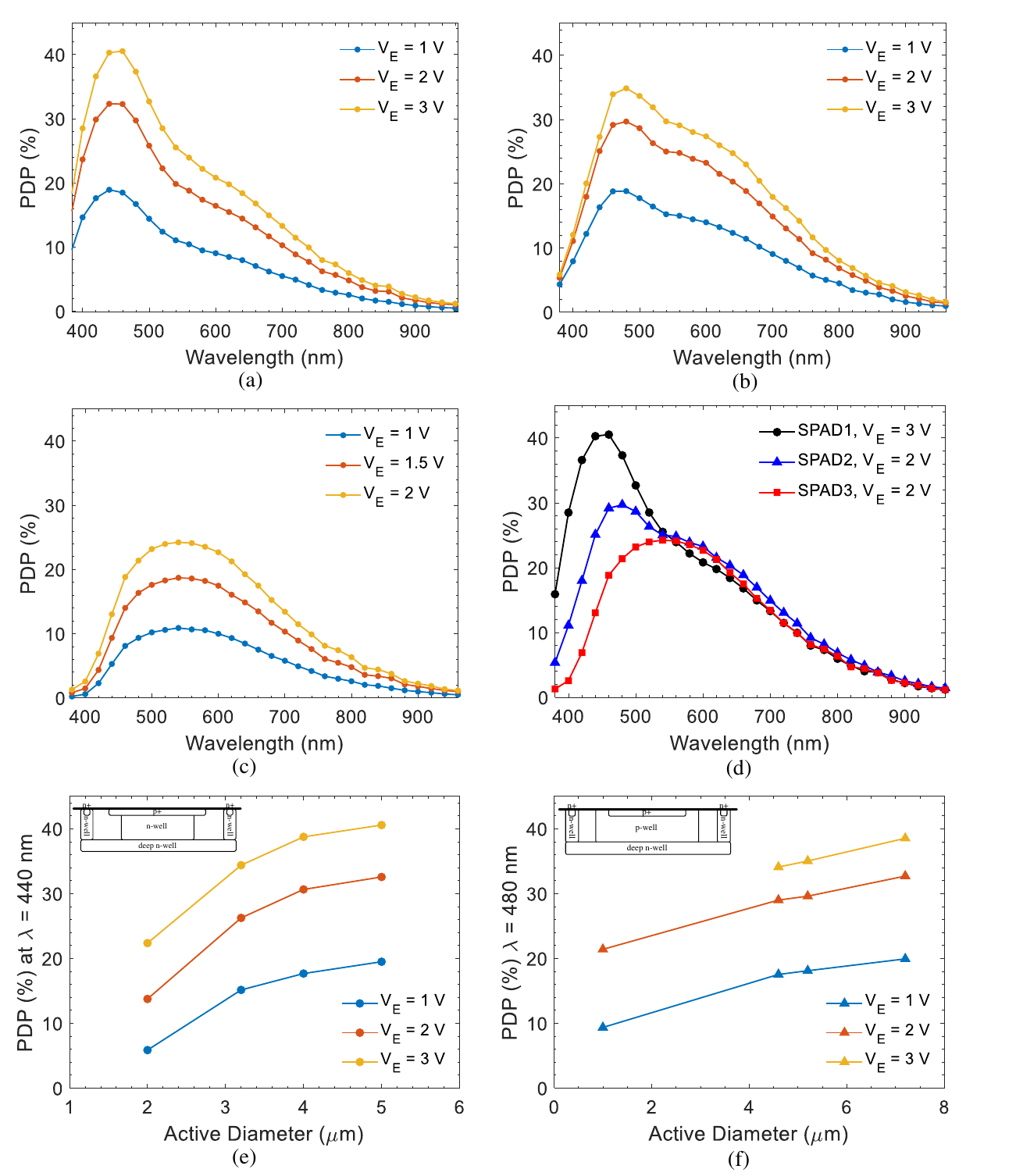}
\caption{ Measured PDP as a function of wavelength for SPAD1 (a), SPAD2 (b), and SPAD3 (c) from 380 nm to 960 nm with the step of 20 nm. (d) PDP comparison as a function of wavelength for the SPAD family. The main reason for showing the PDP performance varying excess bias voltages is that SPAD3 exhibits extremely high crosstalk. Therefore, the PDP measurements of SPAD2 and SPAD3 are compared at an excess bias voltage of 2\,V. (e) PDP at $\lambda$ = 440\,nm as a function of draw active diameter for the p+/nw SPAD; (f) PDP at  $\lambda$ = 480\,nm as a function of draw active diameter for the pw/dnw SPAD. The measured PDP is corrected after removing the afterpulsing and crosstalk. }\label{pdp}
\end{figure}

\subsection{Photon Detection Probability}
Fig.~\ref{pdp} shows the measured PDP from 380\,nm to 960\,nm with a step of 20\,nm for all the SPADs\cite{gramuglia2021engineering}. For the shallow junction, SPAD1 achieves a peak PDP of 40.6\% at 440\,nm, and 1.4\% at 940\,nm at 3\,V excess bias voltage. SPAD2 shows a peak PDP of 29.5\% at 480nm and 1.6\% at 940\,nm at 2\,V excess bias voltage. Compared to SPAD2, the miniaturized SPAD3 has a lower PDP at the wavelength of around 450\,nm. The penetration depth for wavelengths between 400\,nm and 450\,nm is quite close to the surface. When the drawn active diameter of the SPAD is only 1\,$\mu$m, and the p+ layer is even smaller, the electron-hole pair generated close to the surface is thus harder to drift to the avalanche region by the gradual doping. Additionally, side effects can further impact this process. SPAD3 achieves a peak PDP of 9.4\% at 1\,V, and 21.4\% at 2\,V excess bias voltage at 540\,nm. Fig.~\ref{pdp}(e)-(f) shows the PDP as a function of drawn active diameter of the SPAD structures. During the pitch scaling, same width of guard ring is used. We can see that the PDP is decreasing with a smaller drawn active diameter, likely due to the border effects.

\subsection{Crosstalk Probability}
\begin{figure}[ht!]
\centering\includegraphics[width=13cm]{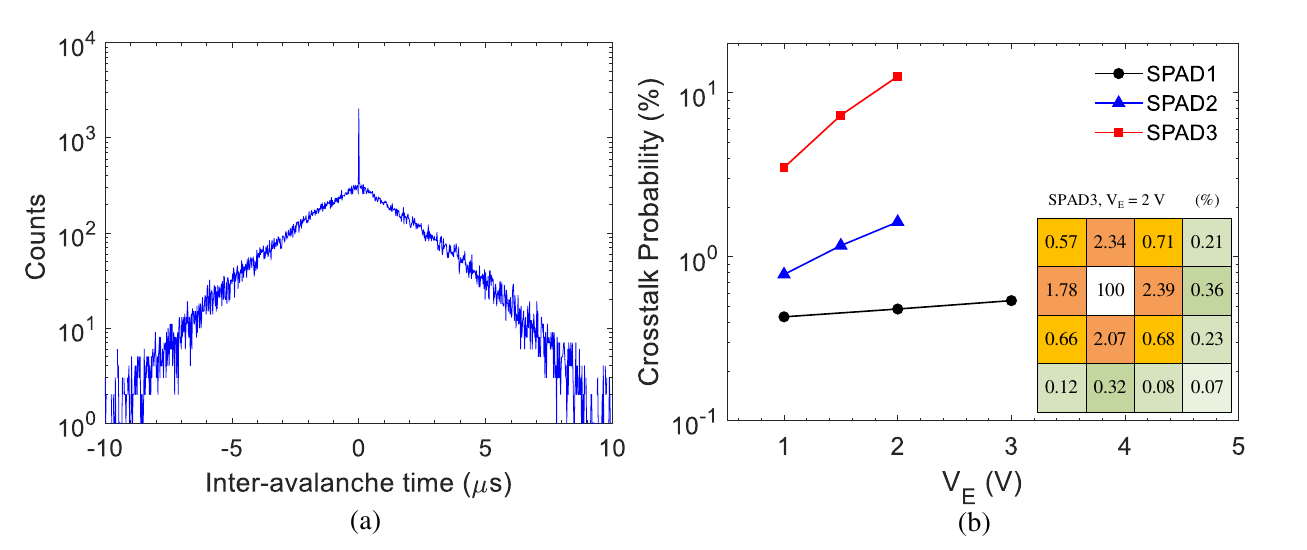}
\caption{ (a) Inter-avalanche time histogram for two adjacent pixels in SPAD3. (b) Crosstalk probability as a function of excess bias voltage of the SPAD family. The inset shows the crosstalk map of SPAD3 at VE = 2\,V.The overall crosstalk for a SPAD is then determined by summing the crosstalk contributions from all neighboring SPADs. }\label{xtalk}
\end{figure}

The crosstalk probability is measured by inter-avalanche time histograms, as in \cite{morimoto2020high}. Fig.~\ref{xtalk}(a) shows the histogram of measured inter-avalanche time for two adjacent pixels under low light condition in SPAD3. Fig.~\ref{xtalk}(b) shows the crosstalk probability as a function of the excess bias voltage for the entire SPAD family. The inset shows the crosstalk map of SPAD3 at 2\,V excess bias voltage. SPAD1 shows the lowest crosstalk probability of 0.5\% at 3\,V excess bias voltage. With similar drawn active diameter while smaller pitch, SPAD2 shows a higher crosstalk probability of 1.6\% at 2\,V excess bias voltage. Degradation of crosstalk is observed in SPAD3. With the guard-ring sharing technique, it shows a crosstalk of 3.5\% at 1\,V, and 12.6\% at 2\,V excess bias voltage.

\begin{figure}[ht!]
\centering\includegraphics[width=13cm]{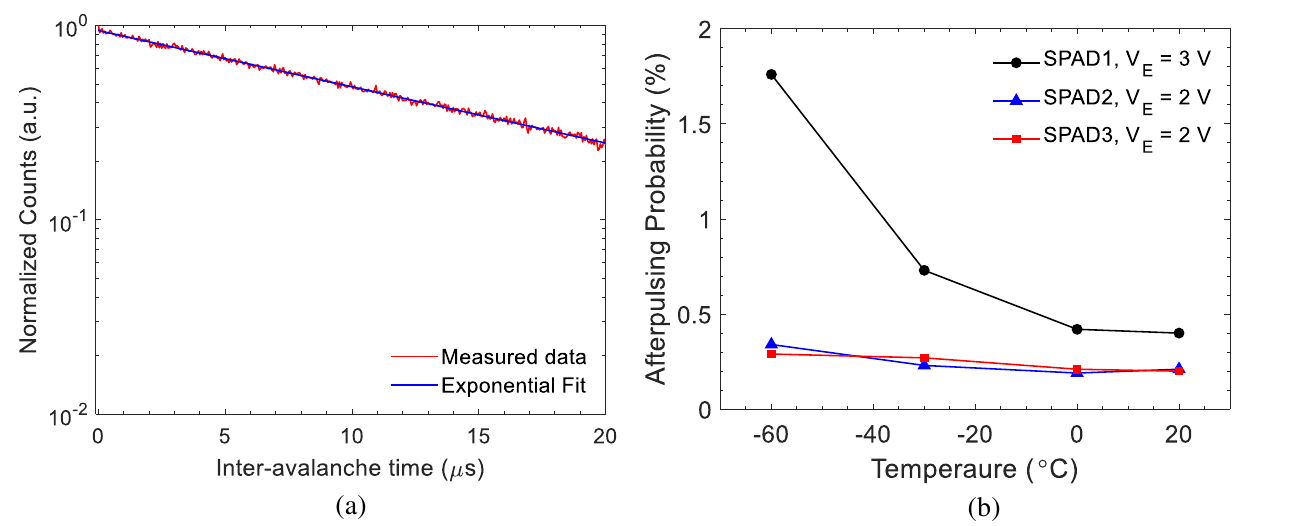}
\caption{ (a) Afterpulsing probability: measured inter-avalanche time histogram at
room temperature (b) Afterpulsing probability as a function of temperature at 3\,V excess bias voltage for SPAD1, and 2\,V excess bias voltage for SPAD2 and SPAD3, respectively. The dead time of each SPAD is 5\,ns.  }\label{app}
\end{figure}

\subsection{Afterpulsing Probability}
When carriers become trapped in 'deep' levels caused by defects in the silicon process, their release and the subsequent avalanche result in secondary pulses known as afterpulses. The release time of these carriers can range from nanoseconds to a few microseconds. Typically, the dead time affects the afterpulsing probability (APP), with a smaller dead time potentially increasing the APP. Afterpulsing probability was measured by the inter-avalanche histogram method for a single pixel under low light condition, as shown in Fig.~\ref{app}(a).
Fig.~\ref{app}(b) shows afterpulsing probability as a function of temperature. The corresponding afterpulsing probability of the SPAD family is 0.4\%, 0.21\%, and 0.2\% at room temperature at 3\,V excess bias voltage for SPAD1, and 2\,V excess bias voltage for SPAD2 and SPAD3, respectively.

\subsection{Timing Jitter}
\begin{figure}[ht!]
\centering\includegraphics[width=8cm]{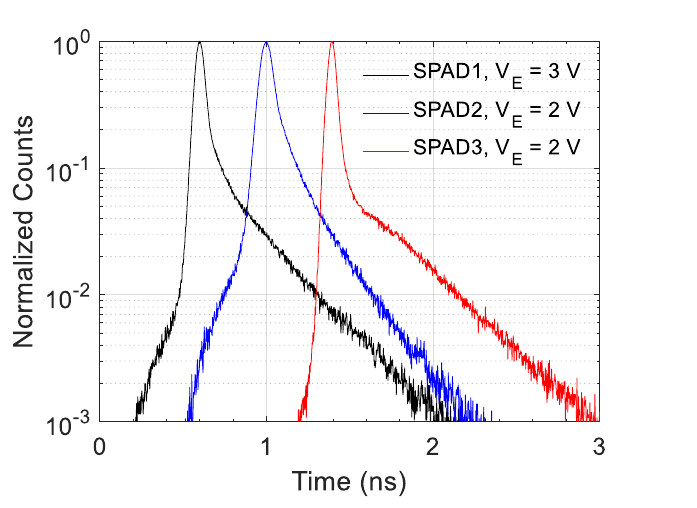}
\caption{ Timing jitter histogram of the SPAD family with an 850\,nm laser.}\label{timing}
\end{figure}

The measured histogram of timing jitter with an 850\,nm pulsed laser emitting 32\,ps pulses is shown in Fig.~\ref{timing}. When applying a quadratic correction to the laser pulse width, a timing jitter of 55\,ps (FWHM) is achieved at 3\,V excess bias in SPAD1. With the pitch scaling down, the timing jitter of deep junction SPAD decreases from 84\,ps to 47\,ps FWHM at 2\,V excess bias voltage. The FW1\%M of the SPADs are calculated to be 832\,ps, 889\,ps, 857\,ps, respectively.

\section{State-of-the-art Comparison}
Table 1 shows the overall performance of the SiGe BiCMOS SPAD family and the state-of-the-art comparison of the miniaturized SPAD arrays. A sub-15V breakdown voltage was achieved, which can help decrease power consumption. A pitch of 1.8\,$\mu$m was demonstrated using guard-ring sharing, the smallest ever reported for a SPAD in any technology. Thanks to the high speed of the process, the proposed SPADs also achieve the best timing jitter at lower excess bias voltage. The crosstalk is relatively high due to the aggressive design of the gap between the neighboring SPADs in guard-ring sharing technique and higher excess bias voltages.

\section{Conclusion}
The demonstrated and characterized SPAD family is, to the best of our knowledge, the world’s first SPAD design in a SiGe BiCMOS process. The 1.8\,$\mu$m pitch was achieved using guard-ring sharing techniques, while a jitter of 47\,ps FWHM was reached. 
The development of silicon SPAD in a SiGe process paves the way to design Ge-on-Si SPAD in this process for SWIR applications, where implementation of the Germanium layer on silicon SPADs will be easier, compared to a standard CMOS platform. The DCR of SPAD decreases from 36.9\,cps at 25\,ºC to 4.9\,cps at -60\,ºC; and could be further improved at deep cryogenic temperatures, where it could be utilized for high-speed communications to quantum processors.

\begin{table}[htbp]
%\centering
\caption{Performance summary and comparison with small pitch SPAD sensor}
\begin{tabular}{   p{1.6cm} p{0.9cm} p{0.9cm} p{0.9cm} p{0.9cm} p{0.9cm} p{0.9cm} p{0.9cm} p{0.9cm} p{0.9cm} }
\hline
Parameter & IEDM 2010\cite{henderson20103} & IISW 2017\cite{ziyang20173mum} & OE 2020\cite{morimoto2020high} & IEDM 2022\cite{shimada2022spad} &  IISW 2023\cite{jogi2023} & This Work &  &  &  \\ 
\hline
Technology & FSI 90nm & FSI 130nm & FSI 180nm & BSI 90nm &  BSI 90nm  &  SiGe 130nm &  &  & \\
Pitch ($\mu$m) & 5 & 3 & 2.25 & 2.5 & 3.06 & 11.3 & 8.3 & 1.8 & 1.8 \\
Array size & 3$\times$3 & 4$\times$4 & 4$\times$4 & N/D & 160$\times$264 & 4$\times$4 & 4$\times$4 & 4$\times$4 & 4$\times$4 \\
Fill Factor (\%) & 12.5 & 14 & 19.5 & N/D & N/D & 15.3 & 30.9 & 24.2 & 24.2\\
VBD (V) & 10.3 & 15.8 & 32.3 & 18 & 20.9 & 12.7 & 10.8 & 11.0 & 11.0 \\
VE (V) & 0.6 & 3.2 & 4 & 3 & 3 & 3 & 2 & 1 & 2\\
DCR (cps) & 250 & 150$^\textit{a}$ & 751 & 173 & 15.8 & 1598 & 5453 & 36.9 & 601 \\
Peak PDP (\%) & 36 & 15 & 10.3 & N/D & N/D & 40.6 & 29.5 & 9.4 & 21.4\\
Peak PDE (\%) & 4.5 & 2.1 & 2.0 & 76.1 & 57 & 6.2 & 9.1 & 2.3 & 5.2\\
Crosstalk (\%) & <0.1 & <0.2$^\textit{a}$ & 2.97 & 1.0 & <0.4 & 0.5 & 1.6 & 3.5 & 12.6 \\
Afterpulsing Probability (\%) & N/D & 0.18$^\textit{a}$ & <0.2 & <0.1 & N/D & 0.4 & 0.21 & <0.2 & 0.2 \\
Dead time (ns) & N/D & 10 & 10 & 6.0 & N/D & 5.0 & 5.0 & 5.0 & 5.0 \\
Timing (ps) FWHM & 107 & 176 & 72 & 214 & N/D & 55 & 84 & 49 & 47 \\
\hline
\end{tabular}
  \label{tab:shape-functions}
$^\textit{a}$The data was measured at 1\,V excess bias voltage.
\end{table}

\begin{backmatter}

%\bmsection{Funding}
%need check.

\bmsection{Acknowledgments}
The authors would like to thank Claudio Bruschini of EPFL, Myung-Jae Lee of KIST for valuable discussions. 

\bmsection{Disclosures}
The authors declare no conflicts of interest. For the sake of transparency, the authors would like to disclose that Edoardo Charbon is co-founder of Pi Imaging Technology and NovoViz. Both companies have not been involved with the paper drafting, and at the time of writing have no
commercial interests related to this article.

\bmsection{Data availability}
Data underlying the results presented in this paper are not publicly available at this time but may be obtained from the authors upon reasonable request.
\end{backmatter}

%%%%%%%%%%%%%%%%%%%%%%% References %%%%%%%%%%%%%%%%%%%%%%%%%

%%%%%%%%%% If using BibTeX:
%\bibliography{reference}

\end{document}